\documentclass[a4paper]{article} 
\usepackage{spconf,amsmath,epsfig}
\usepackage{url}

\title{Towards international E-stat for monitoring the socio-economic
activities across the globe}
\name{Aki-Hiro Sato, Ken Umeno
}
\address{Department of Applied Mathematics and Physics,
Graduate School of Informatics, Kyoto University} % Leave this empty for the paper proposal phase!

\begin{document}

% Change page numbering style *only* for the paper proposal phase!
\pagestyle{plain}
\pagenumbering{arabic}

%\ninept
%
\maketitle
\begin{abstract} \em
We investigate relationship between annual electric power
consumption per capita and gross domestic production (GDP) per capita
for 131 countries. We found that the relationship can be fitted with
a power-law function. We examine the relationship for 47 prefectures 
in Japan. Furthermore, we investigate values of annual electric power production 
reported by four international organizations. We collected the data from
U.S. Energy Information Administration (EIA), Statistics by
International Energy Agency (IEA), OECD Factbook (Economic,
Environmental and Social Statistics), and United Nations (UN) Energy
Statistics Yearbook. We found that the data structure, values, and unit
depend on the organizations. This implies that it is further necessary
to establish data standards and an organization to collect, store, and
distribute the data on socio-economic systems. 
\end{abstract}
\section{Introduction}
Sustainability is the important issue in the current world. The fundamental
idea of sustainability was proposed by R. Backminster Fuller in
\textit{Operating Manual for Spaceship Earth}~\cite{Fuller:1967}. He
proposed the concept that Earth is similar to a spaceship flying through
space. The spaceship has a finite amount of resources and the resources
cannot be resupplied. 

More recently, Brundtland Commission proposed the concept of sustainable
development in 1987. It contains two key concepts; ``needs'' and
``limited resources'' in developing countries. After that, the concept of
sustainable production emerged at the United Nations Conference on
Environment and Development in 1992. The conference concluded that
especially in industrialized countries, the major cause for the continued
deterioration of the global environment is the unsustainable pattern of
consumption and production.

\begin{figure}[!hbt]
\includegraphics[scale=0.32]{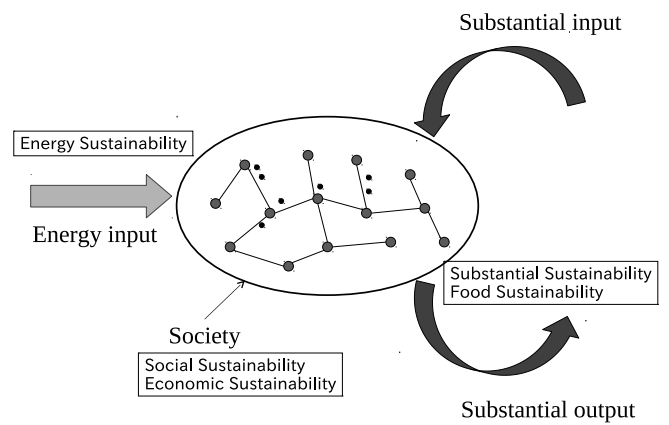}
\caption{The conceptual illustration of society consisting of many
 elements.}
\label{fig:concept}
\end{figure}

Fig. \ref{fig:concept} shows a conceptual illustration of a human
society. The energy injection and substantial inflow/outflow are
mandatory in order to maintain our society. Human resources and social
organizations should be resupplied from the next generation to maintain
our socio-economic systems. Therefore, sustainability of our society may be
classified into several categories: Energy sustainability, substantial
sustainability, food sustaintability, economic sustainability, social
sustainability, and so forth.

Veleva and Ellenbecker proposed framework and methodology to measure
sustainability~\cite{Lowell,Veleva}. Their framework is based on six main
aspects of sustainable production:
\begin{itemize}
\item energy and material use (resources)
\item natural environment (sinks)
\item social justice and community development
\item economic performance
\item workers
\item products
\end{itemize}
To implement their framework in actual situations, we need data on human activities.

Information and Communications Technology (ICT) is expected to contribute for
constructing sustainable society for the next generation. In principle, ICT 
enables us to communicate with one another via computer networks. In the
computer network, large amounts of data on human activities can be
accumulated. These circumstances prepare emergence of data-centric
social sciences at this time. This has also a potential for
reconstructing our social structure from data. 

Our society consists of six billion individuals and various types of
mechanical and electrical equipments. Each element has the associated
geometric information, several properties and states. According to
M.F. Goodchild~\cite{Goodchild}, every human 
is able to act as an intelligent sensor, and in that sense, the earth's
surface is currently occupied by more than six billion sensors. 
We can extract information from an amount of data, construct knowledge
from lots of information, and hopefully establish wisdom from several
pieces of knowledge. Specifically, researchers in the fields of sociology,
economics, informatics, and physics are focusing on these frontiers and
have launched data-centric social sciences in order to understand
the complexity of socio-economic systems. 

D. Weinberger suggests a potential and a limitation of big-data to
predict and understand our society~\cite{Weinberger}. He focuses on an
European large-scale research project called FuturICT
(\url{http://www.futurict.eu}). In this big research project,
large-scale sensor networks,  data analysis platforms, and computer
simulation environments will be constructed. Data-centric investigation in
socio-econo-techno-environmental sciences will be established. This
project is expected to provide us with technological, scientific, and
commercial outcomes to our society.

Measuring properties and states of social elements provides large
amounts of data on socio-economic activities. The number of elements
consisting of our society is enormous and information generated from our
society exceeds our individual's cognitive capacity. In fact, it is 
difficult to grapes states of our social environment, thus it is
necessary to understand state of our society from more precisely and
accurately in order to construct sustainable community.

In this article, we focus on data on electrical power consumption as
energy consumption. The energy production and consumption is one of
useful quantities to measure socio-economic activity. The socio-economic
systems are constructed by many mechanical and electronic equipments
driven by electricity or oils. Therefore, it is reasonable to infer that
socio-economic activities are proportional to gross energy consumption in
a society. The relationship between annual energy consumption and
annual gross domestic product (GDP) has been largely studied in the
context of designing efficient energy conservation policies. The
pioneering study by Kraft and Kraft~\cite{Kraft} reported that causality
runs from GNP to energy consumption using data on gross energy inputs
and gross national product (GNP) for the USA. More recently Narayan et
al. study Granger causality between electricity consumption and real GDP
for 93 countries~\cite{Narayan}. They reported that in the six most industrialized
nations increasing electricity consumption may reduce GDP.

However, a relationship of individual activities between electrical
power consumption and economic productivity is not clarified. One of 
aims in this article is to elucidate the relationship between 
electrical power consumption per capita and GDP per capita. 

Another aim in this article is to show data inconsistency among
organizations which report energy statistics. We further propose the
necessity of data management to understand our social states more
accurately. To construct rigorous database of socio-economic systems, we
need to consider both rules of data collection and roles of organizations. 

This article is organized as follows. In Sec. \ref{sec:relation} we show the
relationship between annual electrical power consumption per capita and
GDP per capita. Sec. \ref{sec:inconsistency} shows
an example of the data inconsistency in energy consumption among
organizations reporting energy statistics. In Sec. \ref{sec:power-data}
we propose the data integrity for electric power and an authentication
of real unified data. Sec. \ref{sec:conclusion} is devoted to conclusions.

\section{Relationship between energy consumption and socio-economic activity}
\label{sec:relation}
Energy production and consumption are deeply related to human activities
in our society. This fact can be partially confirmed from the
relationship between annual electrical power consumption per capita and 
annual GDP per capita. 

\subsection{Relationship for 131 countries}
The table \ref{tab:World} shows the annual electrical power consumption per
capita and GDP per capita in 2009 for 131 countries~\footnote{The 
countries included in the data are ALB (Albania), DZA (Algeria), AGO
(Angola), ARG (Argentina), ARM (Armenia), AUS (Australia), AUT
(Austria), AZE (Azerbaijan), BHR (Bahrain), BGD (Bangladesh), BLR
(Belarus), BEL (Belgium), BEN (Benin), 
BOL (Bolivia), BIH (Bosnia and Herzegovina), BWA (Botswana), BRA
(Brazil), BRN (Brunei Darussalam), BGR (Bulgaria), KHM (Cambodia), CMR
(Cameroon), CAN (Canada), CHL (Chile), CHN (China), COL (Colombia), COD
(Congo  Dem. Rep.), COG (Congo  Rep.), CRI (Costa Rica), CIV (Cote
d'Ivoire), HRV (Croatia), CYP (Cyprus), CZE (Czech Republic), DNK
(Denmark), DOM (Dominican Republic), ECU (Ecuador), EGY (Egypt  Arab
Rep.), SLV (El Salvador), ERI (Eritrea), EST (Estonia), ETH (Ethiopia),
FIN (Finland), FRA (France), GAB (Gabon), GEO (Georgia), DEU (Germany),
GHA (Ghana), GRC (Greece), GTM (Guatemala), HTI (Haiti), HND (Honduras),
HKG (Hong Kong SAR  China), HUN (Hungary), ISL (Iceland), IND (India),
IDN (Indonesia), IRN (Iran Islamic Rep.), IRQ (Iraq), IRL (Ireland), ISR
(Israel), ITA (Italy), JAM (Jamaica), JPN (Japan), JOR (Jordan), KAZ
(Kazakhstan), KEN (Kenya), KOR (Korea  Rep.), KWT (Kuwait), KGZ (Kyrgyz
Republic), LVA (Latvia), LBN (Lebanon), LBY (Libya), LTU (Lithuania),
LUX (Luxembourg), MKD (Macedonia  FYR), MYS (Malaysia), MLT (Malta), MEX
(Mexico), MDA (Moldova), MNG (Mongolia), MAR (Morocco), MOZ
(Mozambique), NAM (Namibia), NPL (Nepal), NLD (Netherlands), NZL (New
Zealand), NIC (Nicaragua), NGA (Nigeria), NOR (Norway), OMN (Oman), PAK
(Pakistan), PAN (Panama), PRY (Paraguay), PER (Peru), PHL (Philippines),
POL (Poland), PRT (Portugal), QAT (Qatar), ROU (Romania), RUS (Russian
Federation), SAU (Saudi Arabia), SEN (Senegal), SRB (Serbia), SGP
(Singapore), SVK (Slovak Republic), SVN (Slovenia), ZAF (South Africa),
ESP (Spain), LKA (Sri Lanka), SDN (Sudan), SWE (Sweden), CHE
(Switzerland), SYR (Syrian Arab Republic), TJK (Tajikistan), TZA
(Tanzania), THA (Thailand), TGO (Togo), TTO (Trinidad and Tobago), TUN
(Tunisia), TUR (Turkey), TKM (Turkmenistan), UKR (Ukraine), ARE (United
Arab Emirates), GBR (United Kingdom), USA (United States), URY
(Uruguay), UZB (Uzbekistan), VEN (Venezuela RB), VNM (Vietnam), YEM
(Yemen Rep.), ZMB (Zambia), and ZWE (Zimbabwe).}. The data were downloaded
from the databank of the world bank (\url{http://data.worldbank.org}). 

\begin{figure}[!hbt]
\includegraphics[scale=0.65]{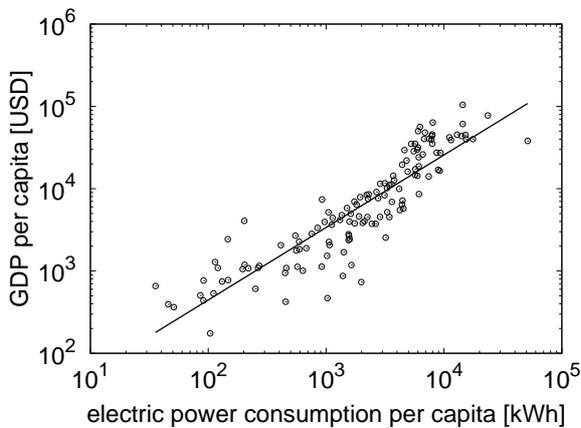}
\caption{The double logarithmic plots between annual electric power
 consumption per capita and GDP per capita. A solid line represents
 a fitting curve.}
\label{fig:energy}
\end{figure}

The figure \ref{fig:energy} shows double logarithmic scatter plots  
between annual electric power consumption per capita and GDP
per capita. From this graph, it is found that there is a monotonically
increasing tendency of GDP per capita according to the annual electric
power consumption per capita. This means that industrialized countries
use annual electrical power more than developing countries. The annual
electrical power consumption per capita and the GDP per capita show a
positive correlation. We assume that the GDP per capita and the annual
electrical power consumption per capita follows a power-law relation;
\begin{equation}
\log y = a \log x + \log C,
\end{equation}
where $y$ is denoted as the GDP per capita, $x$ the annual electrical
power consumption per capita. We obtain $C=7.6975$ and $a=0.8808$ with
the least-squared method.

\subsection{Relationship for 47 prefectures in Japan}
Furthermore, the annual electric power consumption data on each
prefecture in Japan is available from the agency for natural resources
and energy of Japanese Ministry of Economy, Trade 
and Industry (http://www.enecho.meti.go.jp). The population data on each
prefecture in Japan can be also downloaded from a homepage of Statistics
Bureau of Ministry of International Affairs and Communications
(http://www.stat.go.jp). The data of GDP for each Prefecture in Japan are
provided by Cabinet Office (\url{http://www.esri.cao.go.jp/en/sna/memu.html}).
This is called National Accounts of Japan, which is data aggregated from
each prefecture. These data in 2009 were downloaded from these official
sites. 

\begin{figure}[!hbt]
\includegraphics[scale=0.65]{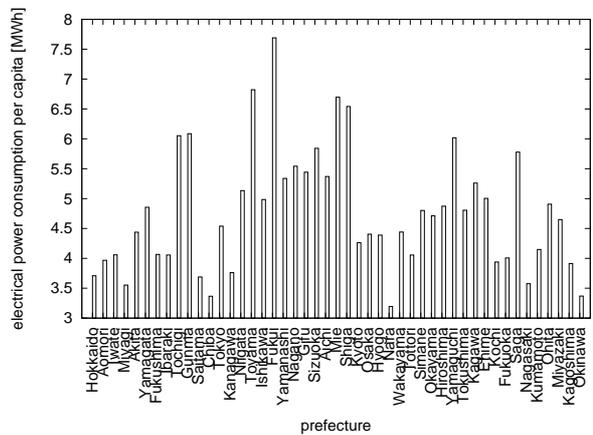}
\caption{The annual power consumption per capita in 2009 for each prefecture.}
\label{fig:electricity}
\end{figure}

\begin{figure}[!hbt]
\includegraphics[scale=0.65]{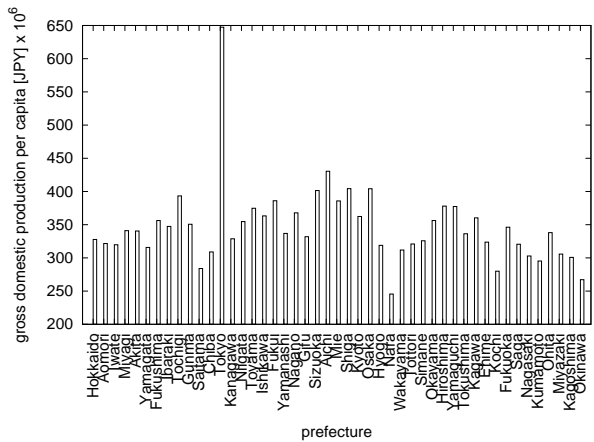}
\caption{The GDP per capita in 2009 for each prefecture.}
\label{fig:gdp}
\end{figure}

The table \ref{tab:JP} shows the annual electrical power consumption per
capita and GDP per capita for 47 prefectures of Japan~\footnote{Japan
consists of 47 prefectures; Hokkaido, Aomori, Iwate, Miyagi, Akita,
Yamagata, Fukushima, Ibaraki, Tochigi, Gunma, Saitama, Chiba, Tokyo,
Kanagawa, Niigata, Toyama, Ishikawa, Fukui, Yamanashi, Nagano, Gifu,
Shizuoka, Aichi, Mie, Shiga, Kyoto, Osaka, Hyogo, Nara, Wakayama,
Tottori, Simane, Okayama, Hiroshima, Yamaguchi, Tokushima, Kagawa,
Ehime, Kochi, Fukuoka, Saga, Nagasaki, Kumamoto, Ohita, Miyazaki,
Kagoshima, and Okinawa.}. Fig. \ref{fig:electricity} shows annual power
consumption per capita for each prefecture. The annual power consumption
per capita in Japan ranges from 3.1 MWh to 7.7 MWh. The highest
consumption per capita is in Fukui, and the lowest is 3.1 in
Nara. Fig. \ref{fig:gdp} shows GDP per capita for each prefecture. The
GDP per capita in Japan ranges from 250 million JPY to 650 million
JPY. The highest is in Tokyo, and the lowest is in Nara.

\begin{figure}[!hbt]
\includegraphics[scale=0.65]{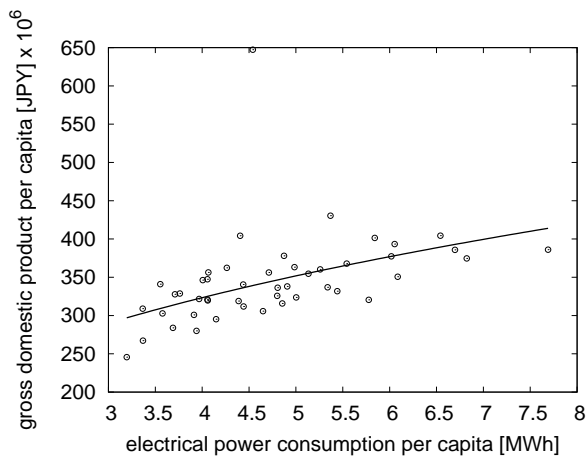}
\caption{The double logarithmic scatter plots between the annual electric power consumption
 per capita and GDP per capita in 2009. A solid
 curve is computed from a power-law fitting.}
\label{fig:JapanFact}
\end{figure}

Fig. \ref{fig:JapanFact} shows the double logarithmic plots
between annual electric power consumption per capita and GDP per capita
of each prefecture in Japan. In the case of Japan, the annual electrical
power consumption per capita ranges from 3 MWh to 8 MWh. In fact, Tokyo is
an outliers, but annual GDP per capita in other prefectures shows
increasing tendency in terms of annual electrical power consumption per
capita. This differences may be created from a difference of human
behavior. Specifically, the situation of Tokyo is different from other
prefectures. This is related to a mechanism of incomes by people in
Tokyo. The number of headquarters of various kinds of organizations such
as companies, institutions, e.t.c. in Tokyo is larger than other
prefectures comparing with its electrical power consumption level.
This implies that domestic productions in Tokyo are given by branches and
factories located in other prefectures. However, a slope between 
the annual electrical power consumption per capita and GDP
per capita is less than the international relationship shown
in Fig. \ref{fig:energy}. We also obtained $C=191.42$ and $a=0.378$ with
the least squared method. The power law exponent $a$ in the whole Japan
is less than those in the whole world.


\section{Example of data inconsistency}
\label{sec:inconsistency}
According to data quality management model, six control dimensions of
data quality are proposed; (1) completeness, (2) accuracy, (3)
duplicates, (4) consistency, (5) integrity, and (6) conformity.
The completeness is that key data items needed are defined in the data
structure. The accuracy implies that the value is consistent with its standard
definition. The duplicates mean that there is only one record in the
tale for a key data. The consistency is that the data in different
tables should be consistent with the rule. The conformity means that
the data should follow the standard format. 

We found that data inconsistency of energy statistics by several
international organizations with the
confidence. Tab. \ref{tab:inconsistency} shows electrical power
productions reported by several international organizations such as 
U.S. Energy Information Administration (EIA),
(\url{http://www.eia.gov/}),
Energy Statistics Yearbook of United Nations Statistics Division (UN)
(\url{http://unstats.un.org/unsd/energy/yearbook/default.htm}), OECD
Factbook 2011-2012: Economic, Environmental and Social Statistics
(\url{http://www.oecd-ilibrary.org/economics/oecd-factbook_18147364}), and
Databank of Worldbank (\url{http://data.worldbank.org}).

\begin{table}[!hbt]
\caption{Electrical power production of several typical countries by
 five international organizations with the confidence in 2009.}
\label{tab:inconsistency}
\centering
\begin{tabular}{ll}
\hline
\hline
United States & electrical power production in 2009 \\
\hline
EIA & 3,950,331,600,000 [kWh] \\
UN & 4,188,214 [GWh] \\
OECD & 4,165.4 [TWh] \\
Worldbank & 4,165,394,000,000 [kWh] \\
\hline
\hline
Germany & electrical power production in 2008 \\
\hline
EIA & 594,685,400,000 [kWh] \\
UN & 637,232 [GWh] \\
OECD & 631.2 [TWh] \\
Worldbank & 631,211,000,000 [kWh] \\
\hline
\hline
Japan & electrical power production in 2009 \\
\hline
EIA & 984,799,000,000 [kWh] \\
UN & 1,047,919 [GWh] \\
OECD & 1,071.3  [TWh] \\
Worldbank & 1,040,983,000,000 [kWh] \\
\hline
\hline
China & electrical power production in 2009 \\
\hline
EIA & 3,445,716,000,000 [kWh] \\
UN & 3,714,950 [GWh] \\
OECD & 3,695.9 [TWh] \\
Worldbank & 3,695,928,000,000 [kWh] \\
\hline
\end{tabular}
\end{table}

According to U.S. EIA , the annual electricity production of the United
States of America is estimated as 3,950,331,600,000 
kWh. UN reports that the annual electricity production of the United
States of America in 2009 is 4,188,214 GWh. However, OECD factbook
reports the annual electricity generation of the United States of
America in 2009 was 4165.4 TWh. UN Energy Statistics Yearbook reports
that the annual electricity generation of the United States of America
in 2009 was 4,188,214 GWh The Worldbank databank reports it was
4,165,394,000,000 [kWh]. The same tendency is confirmed in other
countries.

We found that the unit of the annual electricity generation is not
common. The EIA uses kWh, UN GWh, OECD TWh, and Worldbank kWh. The
values reported by EIA, UN, and OECD are not same. OECD and Worldbank
showed the same value. This means that the statistics are not unique
(The values depend on the associations which report data). Each
association does not seem to communicate with each other and adjust
their reports. 

It is apparent that the inconsistency in the annual electricity
generation is owing to a lack of the coordination agency which is
responsible for standardization and coordination about the annual
electricity generation. Furthermore, frequency of updating the data is
not so fast. Normally the data is delayed for one or two years.

\section{Towards Power Data Integrity}
\label{sec:power-data}
In the case of Japan, Statistics Bureau of the Japanese Ministry of Internal
Affairs and Communication collect macro data of Japan and share them in
a website called E-stat. We suggest that we need the standard of macro
data in socio-economic systems and organize several international
organizations of socio-economic data.

Accurate data of electricity generation is the key ingredient for
building smart community and society with ICT based smart grids for
sustainable development. Thus, the data integrity for electric power and
an authentication of real unified data (standardization of electric
power) is vitally important.  

A simple solution for avoiding the inconsistency is to build an
international institution which is responsible for data consistency
(integrity) in electricity generation and their publications, which is
comparable to building an international root certificate authority (CA)
organization. However, it is not so practical and very hard to build a
new agency. On the other hands, International Telecommunication
Union (ITU) has already issued the recommendation for standardization about data
integrity such as the trusted time source standardization ITU-R TF. 1876~\cite{ITU-R}
for time stamp authority by ITU-R which is important for data integrity.

Furthermore, recent technologies on the smart grids have been intensively
developed. Specifically, automated meter reading (AMR) and smart meters
have been launched~\cite{Lennox,IDC}. ICT may assist automated matching
of electrical power demand and supply. Such an automated matching system
is called a central energy management system, which can be also useful
to measure electrical power generation and consumption in real
time. However, we need to carefully consider balancing consumer privacy
concerns with novel applications of power data to our
society~\cite{McKenna}. To do so, we need several regulations for
organizations handling power data. 

Furthermore, ITU-T Recommendation C. 2~\cite{ITU-T} refers to collection and
publication of official service information for service providers in the field of
telecommunication. Since smart grid services are of an emerging
field, it has not been defined as telecommunication service providers yet.
However, the smart grid service uses telecommunication infrastructure in order to
transmit power data and collect them for commercial purposes. We can
define the smart grid service as one of telecommunication
services. Thus, we will make them observe ITU-T Recommendation C. 2.

Therefore, we propose that ITU plays a leading role for the
standardization for establishing electric power data integrity with time
stamping, which is the main conclusion of our paper.

\section{Conclusion}
\label{sec:conclusion}
We investigated the relationship between electrical power consumption
per capita and GDP per capita for 131 countries using the data reported
by Worldbank. It was found that an increasing tendency
of electrical power consumption per capita in terms of the GDP per
capita. The comparison analysis among countries showed that the relation
can be fitted with the power-law function. Furthermore, we examined the same
relationship for 47 prefectures in Japan. the comparison analysis among
47 prefectures in Japan showed an increasing tendency slower than
comparison among 131 countries. This may indicate that the relationship
between energy consumption and economic activities strongly depends on
life style and social organization in each country.

Moreover, the data inconsistency of international electricity production
was found among EIA, UN, OECD, and Worldbank. We propose an electrical power
data integrity issue and suggest a solution that ITU plays a leading
role for the standardization for building power data integrity and
authentication with the standard time stamping procedure which was
already standardized as ITU-R TF.1876. 

\bibliographystyle{IEEEbib}
\bibliography{aki}

\begin{table}[!hbt]
\caption{The electrical power consumption per capita and GDP per capita
 for 131 countries in 2009. The data source is Databank of Worldbank.}
\label{tab:World}
{\tiny
\begin{tabular}{lll|lll}
\hline
Country & electricity per capita & GDP per capita &
 Country & electricity per capita & GDP per capita \\
         &  [kWh] & [USD] &
         &  [kWh] & [USD] \\
\hline
ALB & 1747.09801006852 & 3795.68886057601 & DZA & 970.982457080035 & 3951.91090053405 \\
AGO & 202.154500255051 & 4068.54853380169 & ARG & 2758.76649642421 & 7665.07344393453 \\
ARM & 1550.41574026922 & 2803.26553438416 & AUS & 11113.3078531503 & 42101.4053099637 \\
AUT & 7944.38915636366 & 45638.0889883881 & AZE & 1620.49918617389 & 4950.29479142375 \\
BHR & 9214.43460803811 & 16517.7718293823 & BGD & 251.628671113669 & 607.764941276909 \\
BLR & 3298.62206794993 & 5182.63040414387 & BEL & 7903.02925218402 & 43849.4146669336 \\
BEN & 91.2602765174753 & 765.553614704077 & BOL & 558.349919951428 & 1774.19520834347 \\
BIH & 2867.01402426903 & 4534.05698137535 & BWA & 1503.34884960254 & 5822.09656455067 \\
BRA & 2206.19652784595 & 8391.66859203938 & BRN & 8661.76496859664 & 27390.050030207 \\
BGR & 4400.84686737777 & 6403.14768611331 & KHM & 130.777842713603 & 744.169961103927 \\
CMR & 271.238195845138 & 1156.81307585744 & CAN & 15471.3844094031 & 39655.7940423275 \\
CHL & 3283.01860308402 & 10178.8907840525 & CHN & 2631.40275503613 & 3748.93449408508 \\
COL & 1047.04853747458 & 5172.91041535865 & COD & 103.855965793197 & 174.505563503457 \\
COG & 146.392676408249 & 2434.00956069012 & CRI & 1812.54206792295 & 6403.5819493475 \\
CIV & 203.462258913761 & 1190.78741491626 & HRV & 3711.89884849853 & 14322.8603402957 \\
CYP & 4620.01351707012 & 29427.9087866487 & CZE & 6114.03754184395 & 18706.8241173825 \\
DNK & 6246.49766118454 & 56329.557003455 & DOM & 1358.19138637595 & 4775.84588349699 \\
ECU & 1115.44552680961 & 3647.69626280873 & EGY & 1548.59357764443 & 2370.7111103535 \\
SLV & 845.071840034361 & 3353.82813810026 & ERI & 51.0004123187181 & 364.204820628504 \\
EST & 5950.28915793895 & 14344.7172737461 & ETH & 45.7581341303567 & 393.697419189245 \\
FIN & 15241.6119437986 & 44889.7515330507 & FRA & 7467.89659221246 & 40477.0644326045 \\
GAB & 922.495489044436 & 7408.65197813071 & GEO & 1585.1999637254 & 2441.01665833044 \\
DEU & 6778.66131414345 & 40275.2507638384 & GHA & 265.106339290279 & 1090.41718145621 \\
GRC & 5540.22242784567 & 28520.9637902007 & GTM & 548.112201674507 & 2688.81322299726 \\
HTI & 35.6844485044516 & 655.930267729905 & HND & 678.262043782198 & 1895.78504481228 \\
HKG & 5924.58272056199 & 29881.8143612784 & HUN & 3773.15380662799 & 12634.551144935 \\
ISL & 51259.1876269627 & 37973.9521896773 & IND & 570.931464603278 & 1126.9451290288 \\
IDN & 590.15352032318 & 2272.7338490967 & IRN & 2237.50863241208 & 4525.94860803344 \\
IRQ & 1068.58104447292 & 2065.92971670014 & IRL & 6033.94258099791 & 50034.1782551533 \\
ISR & 6607.61996366357 & 26032.1635135608 & ITA & 5270.53962591941 & 35073.1580217857 \\
JAM & 1901.61745066034 & 4615.27855835056 & JPN & 7819.18286901394 & 39473.3629058165 \\
JOR & 2111.91885038039 & 4027.05208091558 & KAZ & 4448.07434761939 & 7164.76947131153 \\
KEN & 147.43227111482 & 774.928343540125 & KOR & 8900.28872351673 & 16958.6523899186 \\
KWT & 17609.9635489135 & 40022.6349715782 & KGZ & 1385.76709453309 & 871.218297484413 \\
LVA & 2874.7127282984 & 11475.6923347794 & LBN & 3130.10038146386 & 8321.37074055739 \\
LBY & 4170.1083579887 & 9957.4904063992 & LTU & 3430.79830966481 & 11033.5884564141 \\
LUX & 14423.9558201064 & 104353.691608889 & MKD & 3441.80605600337 & 4528.25473593781 \\
MYS & 3613.53081166873 & 6902.17556058309 & MLT & 4415.55492752258 & 19564.1957457439 \\
MEX & 1942.84079772697 & 7875.82087182458 & MDA & 1018.06061976053 & 1525.53151986014 \\
MNG & 1410.57559228502 & 1690.41698361215 & MAR & 755.566924288161 & 2827.81855367285 \\
MOZ & 453.352209957501 & 423.216539946677 & NAM & 1576.21634929739 & 3983.23755444355 \\
NPL & 90.9531265910214 & 438.288716732201 & NLD & 6895.66391303096 & 47998.2657548238 \\
NZL & 9346.35525279207 & 27196.883167663 & NIC & 459.526148684029 & 1088.16588522573 \\
NGA & 120.507685538337 & 1091.13626599222 & NOR & 23549.6899182103 & 77610.0211604842 \\
OMN & 5723.89119887203 & 17280.0972352585 & PAK & 449.322762669338 & 949.116584867869 \\
PAN & 1735.46268365271 & 6955.74483499095 & PRY & 1055.99401566599 & 2254.05971013169 \\
PER & 1135.71409749057 & 4412.39025697059 & PHL & 593.45873732281 & 1835.63651328558 \\
POL & 3590.8320811579 & 11293.8462174913 & PRT & 4814.58609570183 & 22015.9151038699 \\
QAT & 14420.7690117132 & 61075.0101993575 & ROU & 2266.71746025598 & 7500.3404452926 \\
RUS & 6132.97864843915 & 8615.658757138 & SAU & 7427.21549264699 & 14050.9477408266 \\
SEN & 196.004498274326 & 1054.69424912644 & SRB & 4224.39766544863 & 5484.05356295628 \\
SGP & 7948.91330499639 & 35274.4563548966 & SVK & 4924.71288656274 & 16100.0827063843 \\
SVN & 6103.44129366088 & 24051.0360705709 & ZAF & 4532.02190179876 & 5738.27160868657 \\
ESP & 6006.34817960228 & 31707.3148295322 & LKA & 412.860635696822 & 2057.11324671148 \\
SDN & 114.270085468798 & 1286.14729681583 & SWE & 14141.7204790227 & 43639.5483187755 \\
CHE & 8021.09446861637 & 63568.2446798421 & SYR & 1562.82188661287 & 2691.5976091214 \\
TJK & 1985.29054057042 & 733.874116597373 & TZA & 85.6754152087027 & 505.800108000364 \\
THA & 2044.825059403 & 3835.24818361822 & TGO & 110.812542285405 & 534.760686132888 \\
TTO & 5661.694662098 & 14771.9136872684 & TUN & 1311.25713628875 & 4168.93675437811 \\
TUR & 2297.79685531646 & 8553.74145297029 & TKM & 2445.944230865 & 3745.33228610026 \\
UKR & 3200.46554752862 & 2545.48034107349 & ARE & 11463.6288761121 & 38959.8122211807 \\
GBR & 5691.60580360524 & 35129.4294052485 & USA & 12913.7114285466 & 45191.9382649098 \\
URY & 2670.90152343631 & 9117.37346981325 & UZB & 1635.91117641551 & 1181.84735960066 \\
VEN & 3151.56426155581 & 11605.7983196973 & VNM & 917.570473699506 & 1129.6751504032 \\
YEM & 218.833726405288 & 1077.24014238306 & ZMB & 635.033110532071 & 1006.38818208941 \\
ZWE & 1026.21518436119 & 467.870569902562 &     &	 &                  \\
\hline
\end{tabular}
}
\end{table}

\begin{table}[!hbt]
\caption{The annual electrical power consumption per capita and GDP per capita
 for 47 Japanese prefectures in 2009. The data source is E-stat of Statistics 
Bureau of the Japanese Ministry of Internal Affairs and Communication
 and the agency for natural resources and energy of Japanese Ministry of
 Economy, Trade and Industry.}
\label{tab:JP}
\centering
{\small
\begin{tabular}{lll}
\hline
Prefecture & electricity per capita & GDP per capita \\
           & [kWh] & [JPY] \\
\hline
Hokkaido & 3293.70925294 & 3277879.841 \\
Aomori & 3093.967717138 & 3216647.829 \\
Iwate & 3124.059014264 & 3197689.642 \\
Miyagi & 2563.553477663 & 3409966.886 \\
Akita & 3444.437899561 & 3404829.088 \\
Yamagata & 3194.85469155 & 3157933.553 \\
Fukushima & 2734.06616999 & 3562819.901 \\
Ibaraki & 1434.056489055 & 3473520.352 \\
Tochigi & 2166.053031357 & 3933252.085 \\
Gunma & 2006.085954002 & 3507062.649 \\
Saitama & 1023.688057065 & 2839643.656 \\
Chiba & 1043.365605748 & 3089699.908 \\
Tokyo & 844.539927784 & 6473422.039 \\
Kanagawa & 893.761565285 & 3287204.265 \\
Niigata & 2015.133635749 & 3546678.586 \\
Toyama & 3336.823183376 & 3746759.774 \\
Ishikawa & 2984.984309657 & 3632357.013 \\
Fukui & 3177.690188865 & 3860217.987 \\
Yamanashi & 2025.338294832 & 3368673.3 \\
Nagano & 2435.5424428 & 3678364.184 \\
Gifu & 3265.441580996 & 3318471.016 \\
Shizuoka & 2385.841432395 & 4013965.574 \\
Aichi & 1925.369072591 & 4304688.035 \\
Mie & 2666.69714415 & 3857842.762 \\
Shiga & 2306.542026406 & 4042867.617 \\
Kyoto & 1224.262804925 & 3623406.723 \\
Osaka & 884.405162956 & 4042305.021 \\
Hyogo & 1364.388720592 & 3189360.795 \\
Nara & 1023.195858757 & 2455876.449 \\
Wakayama & 1584.441170046 & 3118555.502 \\
Tottori & 2454.057104635 & 3209074.162 \\
Simane & 2914.8000324 & 3257565.394 \\
Okayama & 2274.71183821 & 3562334.125 \\
Hiroshima & 2004.873312642 & 3780467.769 \\
Yamaguchi & 3006.018701352 & 3773387.526 \\
Tokushima & 2114.805183926 & 3363703.805 \\
Kagawa & 2735.25928602 & 3602834.565 \\
Ehime & 2155.003730528 & 3236972.18 \\
Kochi & 1903.939757271 & 2799865.55 \\
Fukuoka & 1794.00638863 & 3462569.419 \\
Saga & 2725.777264557 & 3205250.268 \\
Nagasaki & 2353.576590784 & 3028234.382 \\
Kumamoto & 2034.14894664 & 2952628.191 \\
Ohita & 2224.907253556 & 3380163.473 \\
Miyazaki & 2134.648527033 & 3056959.617 \\
Kagoshima & 2783.91235667 & 3008137.466 \\
Okinawa & 1933.368651733 & 2672217.582 \\
\hline
\end{tabular}
}
\end{table}

\end{document}